\documentclass[aps,pre,reprint,showpacs,superscriptaddress]{revtex4-1}
\usepackage{graphicx}
\usepackage{subfigure}
\usepackage{color}
\usepackage{amsmath}
\usepackage{lipsum}
\usepackage{amssymb}
\usepackage{epstopdf}
\usepackage[linktocpage,colorlinks=true,linkcolor=blue,citecolor=blue,breaklinks=true]{hyperref}
\usepackage{verbatim}
\usepackage{breakcites}
\usepackage[margin=0.95in]{geometry}
\usepackage{ulem}

\begin{document}

%\preprint{}

\title{Absence of Evidence for the `Ultimate' State of Turbulent Rayleigh-B\'enard Convection}

\author{Charles R. Doering}
\affiliation{Center for the Study of Complex Systems, Department of Mathematics, and Department of Physics, University of Michigan, Ann Arbor, Michigan 48109-1043, USA}
%\author{S. Toppaladoddi}
%\affiliation{University of Oxford, Oxford OX2 6GG, United Kingdom}
%\author{J. S. Wettlaufer}
%\affiliation{Yale University, New Haven, Connecticut 06520-8109, USA}
%\affiliation{University of Oxford, Oxford OX2 6GG, United Kingdom}
%\affiliation{Nordita, Royal Institute of Technology and Stockholm University, SE-10691 Stockholm, Sweden}
\maketitle

There are a number of distinct predictions for the asymptotic behavior of heat transport $Nu$ as the Rayleigh number $Ra \rightarrow \infty$ in thermal turbulence described by the fundamental model of Rayleigh-B\'enard convection \cite{Rayleigh1916}.
One is $Nu = {\cal O}(Ra^{1/3})$ \cite{priestley,malkus,S62,howard} and another is the so-called `ultimate' scaling
$Nu = {\cal O}(Ra^{1/2})$ \cite{spiegel} possibly modified by logarithmic corrections ranging from $Ra^{1/2}/(\log Ra)^{3/2}$
\cite{kraichnan} to $Ra^{1/2}/(\log Ra)^{3}$ \cite{chavanne97}.
%The `classical' $1/3$ scaling theories are generally \cite{priestley,malkus,S62,howard} but not always \cite{kraichnan} uniform in the Prandtl number $Pr$ while the `ultimate' $1/2$ theories all have a leading $Pr^{1/2}$ prefactor \cite{dependence}.

He {\it et al.}~\cite{HFNBA2012PRL} reported measurements (Appendix \ref{appendix}) of $Nu$ for $Ra \in [3 \times 10^{12} ,10^{15}]$ citing them as evidence of transition to the `ultimate' state as characterized by the pre-asymptotic multi-parameter fit in~\cite{GL2011PoF}.
In this comment, without questioning the veracity of the measurements (they have been questioned \cite{doubt}) we show that
%, speaking for themselves,  
the data do not support the claim.

\begin{figure}[h]
\centering
\includegraphics[scale = 0.29]{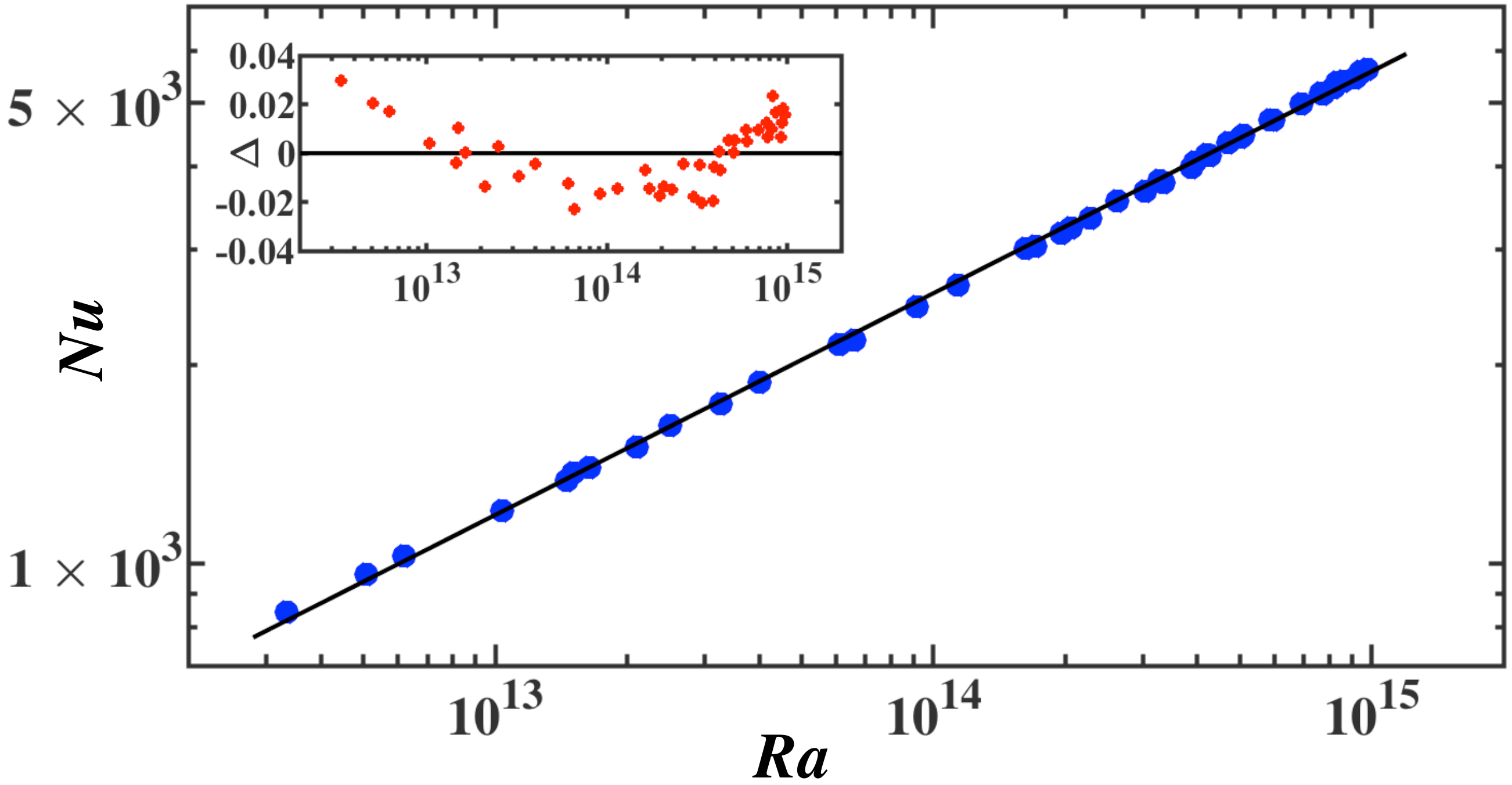} 
%\vspace{-0.2 cm}
\caption{$Nu$ vs.~$Ra$ data from \cite{HFNBA2012PRL} and the power law fit $Nu = 0.0502 \times Ra^{0.336}$.  Inset:
%deviations of the data from the fit, i.e., 
$\Delta \equiv Nu^{data}/Nu^{fit}-1$.}
\label{fig:NuRa1}
%\vspace{-0.5 cm}
\end{figure}

Figure \ref{fig:NuRa1} shows the data with a linear least-squares fit of $\log Nu$ to $\log Ra$ yielding $Nu = 0.0502 \times Ra^{0.336}$.
%reported by He {\it et al.}~
This agrees remarkably with---indeed extends---the $Nu = 0.0508 \times Ra^{1/3}$ fit to experimental data, within about $\pm5\%$, 
%for a comparable Prandtl number
in the overlapping Rayleigh number range $Ra \in [2 \times 10^{11} ,5 \times 10^{13}]$ \cite{UMS}.
%Correspondence with the totally {\it a priori} (i.e., without empirical fit) theoretical prediction
%uniform-in-$Pr$
%$Nu = 0.07 \times Ra^{1/3}$ is impressive \cite{S62}.
% \cite{DG}.

%\vspace{-0.3 cm}

%The apparent precision of
He {\it et al.}'s data, however, suggests more structure than pure power law scaling.
The inset of Figure \ref{fig:NuRa1} shows the $\pm 2.9\%$ ($2\sigma$) deviations from the pure power law fit with a systematic trend that calls for fitting to functional forms capable of capturing the data's convexity.
Data and theories without pure scaling can be compared by examining local slopes $\frac{d\log Nu}{d\log Ra}$.
If data are sufficiently dense then finite difference approximations may be extracted \cite{LS} but the data at hand are not so local slopes can at best be estimated from derivatives of statistically equivalent fits.

\begin{figure}[h]
\centering
%\vspace{0.2cm}
\includegraphics[scale = .41]{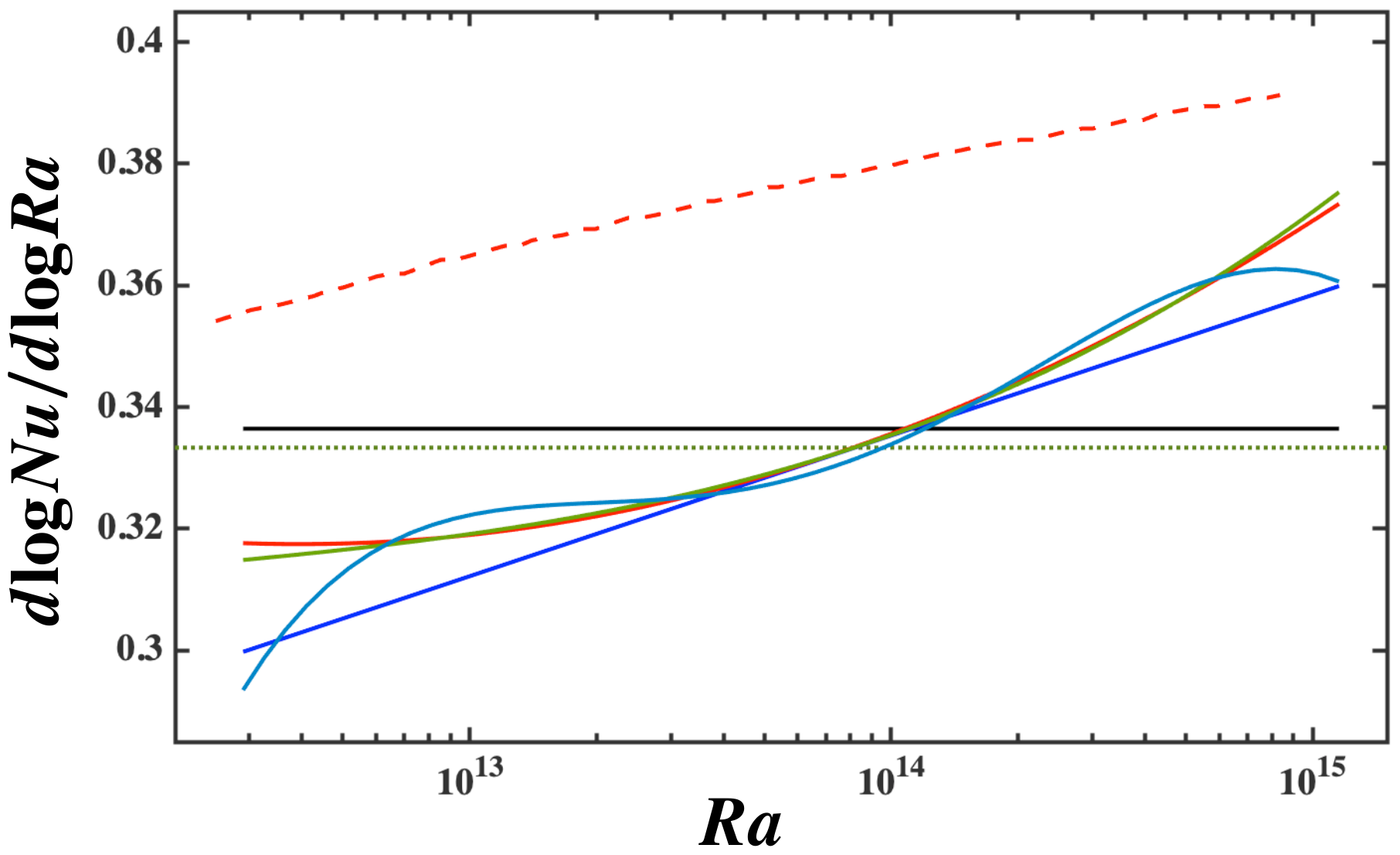}
\caption{Solid lines: local slopes from $1^{st}$ (black), $2^{nd}$ (blue), $3^{rd}$ (red), $4^{th}$ (green) and $5^{th}$ (light blue) order polynomial fits of $\log Nu$ to $\log Ra$.  Dashed line: theoretical pre-asymptotic fit from \cite{GL2011PoF}.  Dotted line: $\frac{d\log Nu}{d\log Ra}=\frac{1}{3}$.
%(Axes same as inset to Figure \ref{fig:NuRa1}.)
}
\label{fig:Fig2}
\vspace{-.2 cm}
\end{figure}

For quadratic, cubic, quartic and quintic polynomial least-squares fits of $\log Nu$ to $\log Ra$, residual deviations are, respectively, $1.19\%$,  $1.09\%$, $1.08\%$ and $1.07\%$ ($2\sigma$) variations with no systematic trends; see Appendix \ref{appendix}. 
Thus each is an equally valid quantitative description of the data, and
%Remaining fluctuations represent irreducible measurement error and uncertainty.
%--------------
%\vspace{0.1cm}
%\centering
%\vspace{0.2cm}
%\includegraphics[scale = .35]{Fig2.pdf}
%\caption{Deviations $\Delta = Nu^{data}/Nu^{fit}-1$ of data from (a) $2^{nd}$, (b) $3^{rd}$, (c) $4^{th}$ and (d) $5^{th}$ order polynomial fits.
% of $\log Nu$ to $\log Ra$.
%(Axes same as inset to Figure \ref{fig:NuRa1}.)
%}
%\label{fig:Fig2}
%\vspace{-.2 cm}
%\end{figure}
%-----------------
%Power law scaling is tested by comparing a theoretical exponent with the slope of the linear fit of $\log Nu$ to $\log Ra$, indistinguishable from $1/3$ for these data.
%Variations in the local slopes of such equivalent fits quantify the essential uncertainty in the data.]
Figure \ref{fig:Fig2} shows local slopes computed from these equivalent fits.

He {\it et al.}~\cite{HFNBA2012PRL} drew a line with $\frac{d\log Nu}{d\log Ra}=0.38$ at the high end of their data citing correspondence with a value from \cite{GL2011PoF} at $Ra = 10^{14}$, but $0.333 < \frac{d\log Nu}{d\log Ra} < 0.336$ for all of the equivalent fits at $Ra = 10^{14}$.
They also reported a transition to $Re \sim Ra^{1/2}$ Reynolds number scaling (necessary but not sufficient for $Nu \sim Ra^{1/2}$ scaling) for $Ra > 5 \times 10^{14}$.
The scaling fit to those data, however, is $Nu = 0.0261 \times Ra^{0.356}$ while local slopes of equivalent fits barely reach $3/8 = 0.375$ (a bound on heat transport dominated by a single horizontal length scale \cite{H63}) at $Ra = 10^{15}$.
But the theoretical slope from \cite{GL2011PoF} is well above $0.39$ there.
%Moreover, their data do not rule out $Nu \sim Ra^{3/8}$, a bound on heat transport dominated by single horizontal length scale \cite{H63}.
%\cite{1015}.
Thus the claim by He {\it et al.}~\cite{HFNBA2012PRL} that their experiment reached the `ultimate' regime
%characterized by heat transport $Nu \sim Ra^{1/2}$ with logarithmic corrections in their experiment
is not justified by their data.
% which also do not rule out $Nu \sim Ra^{3/8}$, a bound on heat transport dominated by single horizontal length scale \cite{H63}.
%Rather, in the context of surrounding results \cite{UMS} and in stark contrast to Kraichnan's theoretical \cite{kraichnan} prediction, the data may be viewed as evidence of robust $Nu \sim Ra^{1/3}$ scaling \cite{Appendix}.
%Rayleigh-B\'enard's `ultimate' regime remains unrealized.

\begin{acknowledgments}
The author thanks S. Toppaladoddi for extracting the $Nu$-$Ra$ data, tabulated in Appendix \ref{appendix}, from \cite{HFNBA2012PRL}.
This work was supported in part by National Science Foundation awards DMS-1515161 and DMS-1813003.
% available in the Supplementary Material.
\end{acknowledgments}

\section{Appendix}
\label{appendix}

\begin{figure}[h]
\centering
\includegraphics[scale = .96]{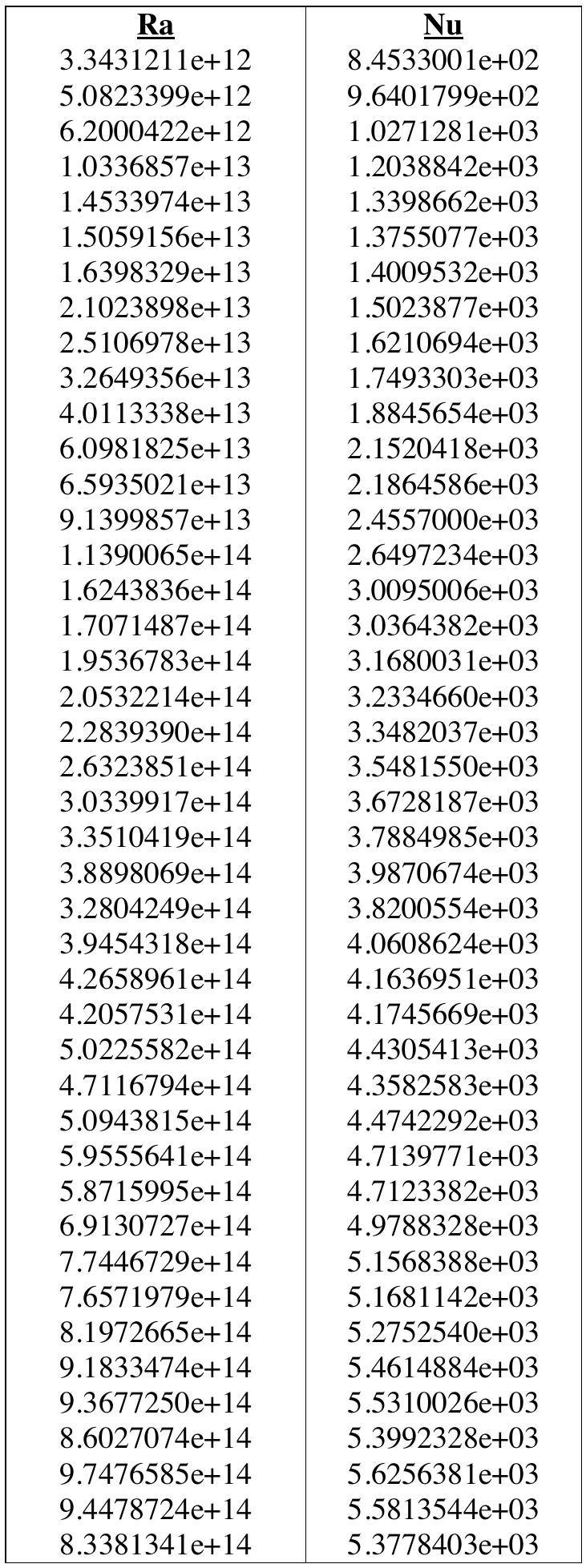} 
%\vspace{-0.2 cm}
\caption{$Nu$ vs.~$Ra$ data from Figure 1(a) of He {\it et al.}~\cite{HFNBA2012PRL} kindly extracted by Srikanth Toppaladoddi.}
%\label{fig:NuRa1}
%\vspace{-0.5 cm}
\end{figure}

\newpage

\begin{figure}[h]
\vspace{1.20 cm}
\centering
\includegraphics[scale = .80]{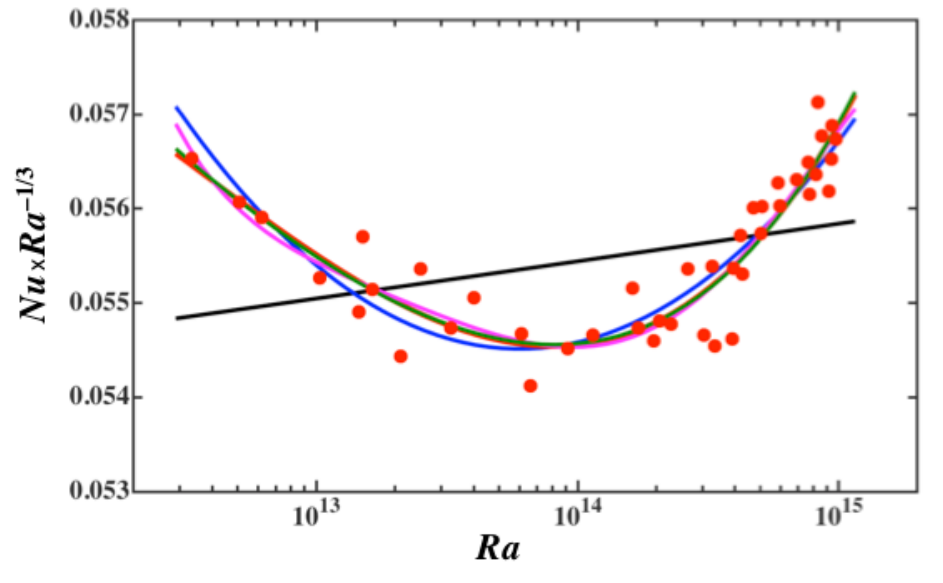} 
%\vspace{-0.2 cm}
\caption{Nusselt vs. Rayleigh data on compensated log-linear plot and least squares polynomial fits of $\log Nu$ to $\log Ra$: black line: linear; blue curve: quadratic, red: cubic, green: quartic, magenta: quintic.}
%\label{fig:NuRa1}
%\vspace{-0.5 cm}
\end{figure}

\begin{figure}[h]
\vspace{.66 cm}
\centering
\includegraphics[scale = .90]{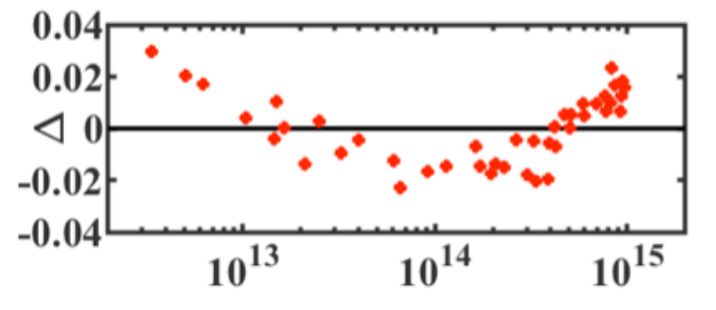} 
%\vspace{-0.2 cm}
\caption{Residuals  ($\Delta \equiv Nu^{data}/Nu^{fit}-1$) of the linear fit of $\log Nu$ to $\log Ra$: $2\sigma = 2.9\%$.)}
%\label{fig:NuRa1}
%\vspace{-0.5 cm}
\end{figure}

\begin{figure}[h]
\vspace{.66 cm}
\centering
\includegraphics[scale = .55]{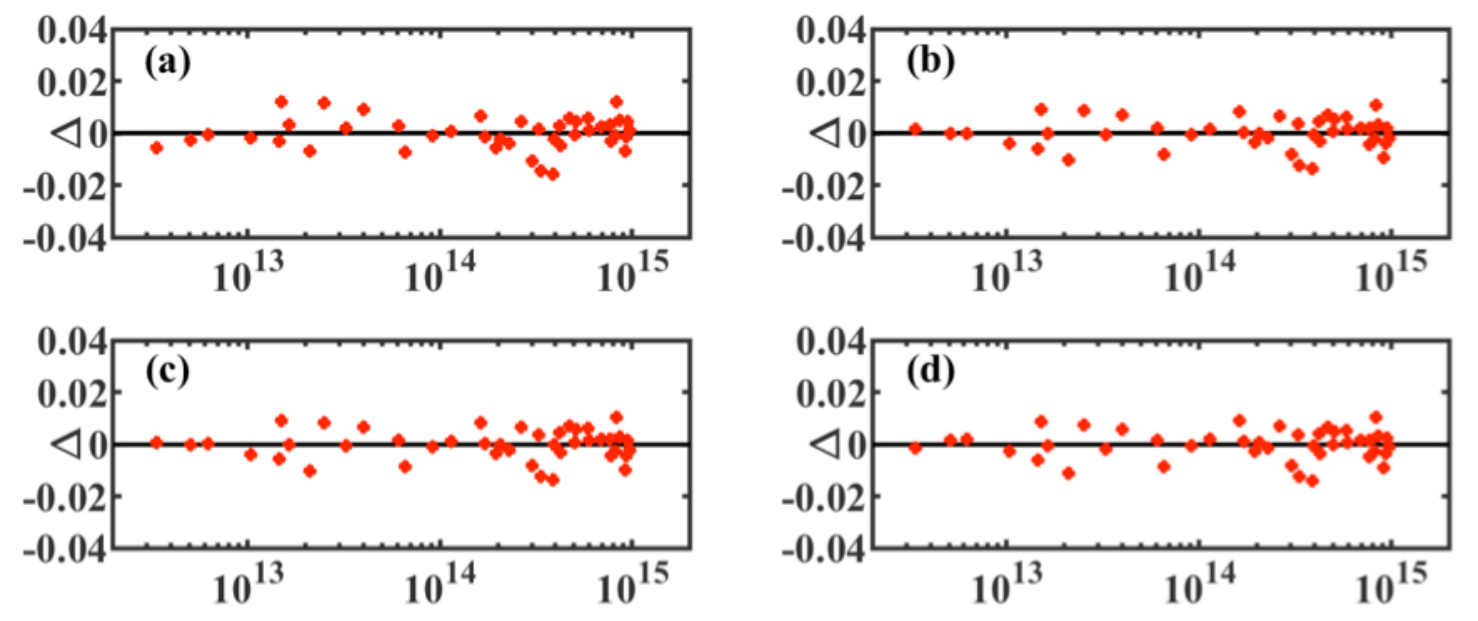} 
%\vspace{-0.2 cm}
\caption{Residuals to polynomial fits of $\log Nu$ to $\log Ra$ ($\Delta \equiv Nu^{data}/Nu^{fit}-1$): (a) quadratic $2\sigma = 1.19\%$, (b) cubic $2\sigma = 1.09\%$, (c) quartic $2\sigma = 1.08\%$, (d) quintic $2\sigma =  1.07\%$. }
%\label{fig:NuRa1}
%\vspace{-0.5 cm}
\end{figure} 

\begin{figure}
\vspace{.86 cm}
\centering
\includegraphics[scale = .80]{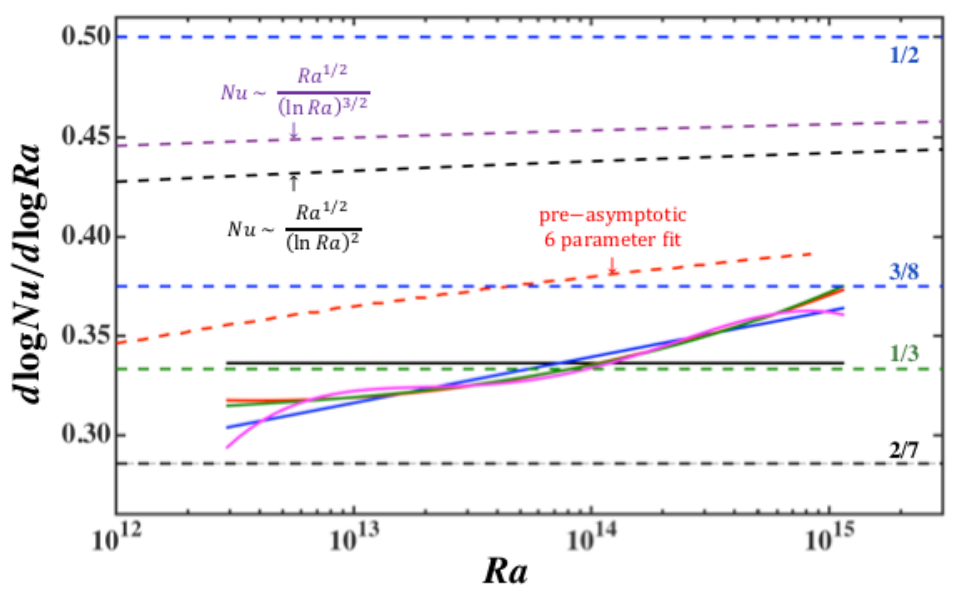} 
%\vspace{-0.2 cm}
\caption{Local slopes (exponents) from data and theories: Horizontal dashed lines indicate pure scaling exponents; Purple and black dashed curves: asymptotic ultimate theories logarithmic corrections; dashed red curve: local slope of multiparameter pre-asymptotic fit \cite{GL2011PoF}.  Black horizontal line: $0.336$ exponent of power law fit to data; solid blue curve: local slope of quadratic equivalent fit to data; solid red curve: local slope of cubic equivalent fit to data; solid green curve: local slope of quartic equivalent fit; solid magenta curve: local slope of quintic fit.}
%\label{fig:NuRa1}
%\vspace{-0.5 cm}
\end{figure}

\begin{figure}
\vspace{.86 cm}
\centering
\includegraphics[scale = .78]{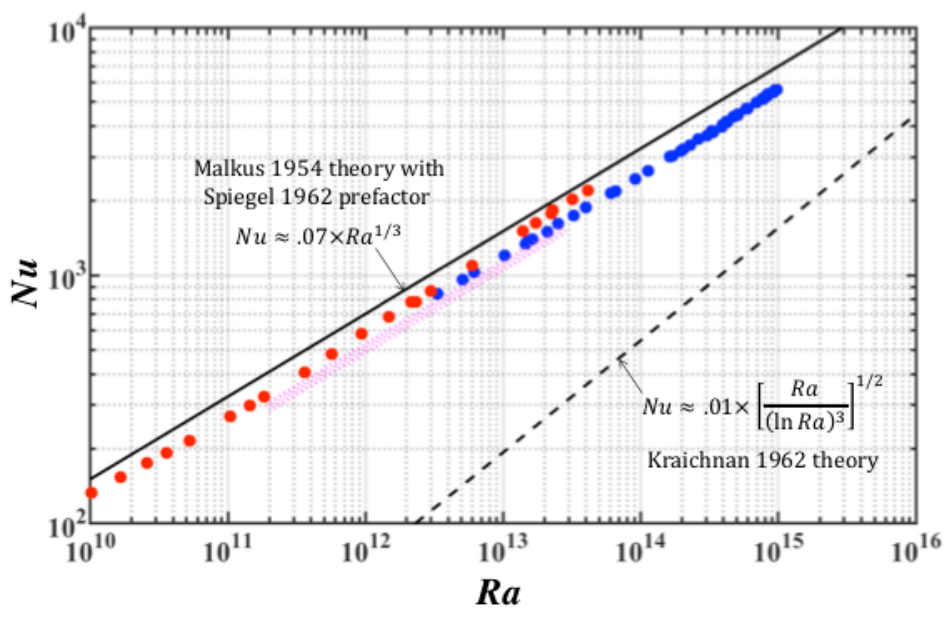} 
%\vspace{-0.2 cm}
\caption{Nusselt vs. Rayleigh data and some asymptotic theories.  Top solid line: Malkus theory \cite{malkus}, uniform in $Pr$,  with {\it a priori} prefactor computed by Spiegel \cite{S62}.   Bottom dashed line: Kraichnan theory \cite{kraichnan} with {\it a priori} prefactor for $Pr = 0.7$. Red large dots: Data \cite{NS} with $0.67 \le Pr \le 1.8$;  magenta small dots: data \cite{UMS} with $0.67 \le Pr \le 2.4$ fit $Nu = .0508 \times Ra^{1/3}$ within $\pm 5\%$; blue large dots: data \cite{HFNBA2012PRL} with $0.79 \le Pr \le 0.86$ and power law fit $Nu = .0502 \times Ra^{0.336}$ within $ \pm 3 \% $ variation.}
%\label{fig:NuRa1}
%\vspace{-0.5 cm}
\end{figure}

\end{document}